\documentstyle[12pt,aasms4]{article}

\begin{document}

\title{Characteristics of MHD Oscillations Observed with MDI}
\author{A. A. Norton\altaffilmark{1} and R. K. Ulrich}
\affil{Department of Physics and Astronomy, University of California, \\
Los Angeles, CA 90095}

\author{R. I. Bush}
\affil{W.W. Hansen Experimental Physics Laboratory, Center for Space Science and Astrophysics, Stanford University, Stanford, CA 94305}

\author{T. D. Tarbell}
\affil{Lockheed Martin Research Laboratory, Palo Alto, CA 94304}

\altaffiltext{1}{Visiting Scholar, Physics Department, Univ.\ of Queensland, Brisbane, QLD 4072 Australia} 

\begin{abstract}
We report on the spatial distribution of magnetogram oscillatory power 
and phase angles between velocity and magnetogram signals 
as observed with the Michelson Doppler Imager.  The dataset is 
151.25\arcsec $\times$ 151.25\arcsec~ containing sunspot from 
Dec 2, 1997 with a temporal sampling interval of 60 seconds and 
spatial sampling of 0.605\arcsec.  Simultaneously observed continuum 
intensity and surface velocity accompany the magnetic information. 
We focus on three frequency regimes:   0.5-1.0, 
3.0-3.5 and 5.5-6.0 mHz corresponding roughly to timescales of 
magnetic evolution, $p$-modes and the 3 minute resonant sunspot oscillation.  
Significant low frequency magnetogram power is found in lower flux pixels, 
100-300 Gauss, in a striking ring with filamentary structure surrounding 
sunspot.  Five minute magnetogram power peaks in 
extended regions of flux 600-800 Gauss.  The 3 minute oscillation 
is observed in sunspot umbra in pixels whose flux measures  
1300-1500 Gauss. Phase angles of approximately -90$^\circ$ between
velocity and magnetic flux in the 3.0-3.5 and 5.5-6.0 mHz regimes 
are found in regions of significant cross amplitude.  
\end{abstract}


\section{Introduction}
The interactions of photospheric magnetic fields 
with motions of solar plasma is described by magnetohydrodynamics (MHD). 
Ionson (1978) recognized that acoustic waves perturbing the base of
magnetic fields can generate MHD waves.   
Significant amounts of literature exist addressing theoretical generation 
and propagation of MHD modes in the solar atmosphere.  Previous 
identification of oscillatory modes in sunspots from velocity and 
intensity measurements is not sufficient since MHD 
modes require simultaneous magnetic field and velocity 
information.    Observational support of solar MHD modes is 
sparse (\cite{ulr96}, \cite{hor97}, \cite{nor97}, \cite{lit98}, 
\cite{rue98}).

Although many MHD waves may exist, two main mechanisms alter 
measured magnetic flux:  1) bending of field lines and 2) 
compression of field lines.  Under simplified conditions, 
the bending mode corresponds to an Alfv\'en wave and a 
compressional mode corresponds to a magnetoacoustic wave.   
The observational tendency to image active regions 
at disk center may have historically hindered Alfv\'en wave detection.
Measurement of $\delta$B due to a change in direction is 
easier to detect in transverse fields than in line-of-sight fields. 
It is easier to detect bending modes at the limb 
and compressional modes at disk center.
%is better viewed at disk center due to decrease in measurement 
%ability of the area element at the limb.  

MHD waves are a prime coronal heating candidate.
Suggested MHD wave dissipation processes are phase-mixing 
(\cite{hey83}) and resonant absorption (\cite{dav87}). 
Detection of MHD oscillations is the first step towards understanding 
the role MHD waves play in atmospheric energy transport.
 
\section{Observations} 

The MDI instrument images the sun at five different wavelengths 
centered around the mid-photospheric Ni-1 6768 \AA~ line. 
Obtaining filtergrams at the five wavelengths samples the line profile. 
The average of the left (LCP) and right circular polarized (RCP)
observed central wavelengths is the Doppler shift.
A longitudinal magnetic flux indicator, uncorrected for observation 
angle, is measured by subtracting 
the observed central wavelengths of RCP from LCP images.
The continuum intensity is measured at a wavelength far from line center.  
We analyze a 151.25\arcsec $\times$ 151.25\arcsec~ sunspot centered 
region in the high resolution field on Dec 2, 1997.  The dataset 
consists of 492 B, I$_c$ and $v$ images, snapshots shown in lower 
panels of Fig 1.  The region is tracked by adjusting heliographic 
coordinates of map centers as a function of time.  The average 
center to limb angle of the sunspot center is 19.77$^\circ$.  
The data is stored as 3-D datacubes, sides of 250 $\times$ 250 
pixels $\times$ 492 minutes. 

\section{Analysis and Results}

Temporal variations from individual pixels are analyzed without 
spatial averaging.  Signals are detrended using a Gaussian filter 
(30 min width) before average values are subtracted.  Power spectra 
of the signals are computed.  3-D datacubes are created 
where the temporal axis has been transformed into frequency.  
Averaging over the selected frequencies, we plot the spatial 
distribution of power for B, I$_c$ and $v$ in Fig 1.
Magnetogram power plotted in Fig 1 is normalized so that the noise, 
as measured in the 7.5-8.0 mHz high frequency band, is unity.  
High frequency power in the strongest magnetic flux regions is 
enhanced by instrumental effects due to reduced intensity 
and broader absorption lines or by increased solar variations.  We 
cannot distinguish between these options but if the high frequency 
power for these pixels is solar, our normalization could be incorrect.  

Averaged magnetogram power spectra are plotted in Fig 2.
Each signal is divided by its standard deviation.  Then 399 
spectra from pixels whose absolute mean measured flux is within 
100-300, 600-800 and 1300-1500 G ranges are averaged and 
plotted in Fig 2.  The ranges are selected because low 
frequency, $p$-mode and 3 minute power peaked herein.

Cross amplitude and phase spectra for the selected frequency regimes are 
found in Fig 3.  To compute the cross spectra, signals are interpolated 
onto a ten second grid and shifted past each other 
in ten second lag increments 
up to a $\pm$15.16 minute period.  The resulting cross covariance function 
is recorded.  Restricting the lag interval to ensure wave train coherence 
time is equivalent to applying a Bartlett window.  The Fourier 
transform of the cross covariance function, the cross spectrum, is computed 
at each position.  Cross amplitude and phase spectra are plotted in the 
left and right columns of Fig 3, respectively. 

\section{Spurious Contributions to Oscillatory Magnetic Signal}
Attention is given to two effects that could mimic an oscillatory magnetic
 signal:
1) misregistration of the LCP and RCP images and ~
2) optical depth changes due to temperature fluctuations
affecting the measurement height where magnetic field gradients, dB/dz, 
are present. To determine whether magnetogram oscillations 
have their origin in LCP-RCP image differences, misregistered datasets were 
simulated. 
The details of this analysis will follow in a subsequent publication.  
Let it suffice herein to note that spurious contributions due to 
misregistered images are most worrisome in quiet regions.  In active 
regions, the suppression of $v$ amplitudes decreases the crosstalk 
between the $v$ and B signals. 

Opacity changes due to 
temperature fluctuations can not be ruled out as a 
source of spurious oscillatory magnetic signal without an 
indicator such as thermal line ratios.     
The analysis of MDI flux estimates as a function of temperature 
will be presented in a subsequent publication.   We acknowledge that
crosstalk into the $\delta$B signal from fluctuations in temperature,
density and other parameters may contribute to magnetogram
oscillations measured herein.  Further research will determine 
what, if any, corrections are necessary to convert magnetogram variation 
into a measure of magnetic field strength variation.

\section{Discussion}

Most significantly, spatial distribution of magnetogram
power is distinctly different in the three frequency regimes.  
Low frequency buffeting and evolution occurs around sunspot and plage. 
Notable filamentary structure seen across the penumbra/quiet 
boundary intimates an advection of flux at low frequencies.   
Magnetogram oscillations on the 5 minute timescale 
are presumably the magnetic response to velocities already
present in the photosphere.  Five minute magnetogram 
oscillations in extended plage regions suggest the plage environment
may more readily convert acoustic waves into MHD waves.  
The 3 minute resonant sunspot oscillation is found in a portion of the umbra 
not associated with the strongest flux, but rather an area bounding 
the darkest part of the umbra.

The uneven spatial distribution of power would cause
an investigation restricting its MHD wave search to the 
strongest flux areas to be unsuccessful.
It is not yet clear what conditions are favorable for the 
generation of measurable MHD oscillations.  Loop termination points or 
strong gradients in magnetic fields might be required.  The 
power distribution and small scale nature of the 
magnetic element does not lend itself to spatial averaging.  
The power spectra shape seen in Fig 2 becomes flatter with 
increasing flux.  Stronger fields   
evolve less, leveling the spectra at low frequency.  The increase of
high frequency power in strong flux spectra is due to reduced intensity
and broader absorption lines or by increased solar variation. The changing
shape of magnetogram spectra should be taken into account
when comparing power from regions of differing B values. 
 
Some positions containing strong magnetic power at 3.0-3.5 mHz in Fig 1
do not show correspondingly high ($v$, $\delta\vert$B$\vert$) cross amplitude
values in Fig 3.  This suggests the presence of different MHD modes or 
crosstalk mechanisms.  The mode with stronger line-of-sight velocity 
variations is more visible in the cross amplitude plot. 

Phase angles of approximately -90$^\circ$ between 
($v$, $\delta\vert$B$\vert$) can be interpreted as $\delta\vert$B$\vert$ 
reaching its maximum a quarter of a cycle before $v$.  Phases of -90$^\circ$ 
dominate regions of significant cross amplitude in the 5 
and 3 minute bands.  This phase relation is not a measure of magnetoacoustic
waves in which $v$ is expected to lead $\delta\vert$B$\vert$.  
Its interpretation is still uncertain, but R\"{u}edi et al (1999) 
demonstrates that $\delta\vert$B$\vert$ leading $v$ a quarter of a cycle 
could be the result of measuring dB/dz during opacity changes.  
It appears that at least three different mechanisms (crosstalk or other) 
dominate at different spatial positions, producing the 
phase structure seen in Fig 3.  
We will not know how the observational bias against detecting
Alfv\'en waves at disk-center affects phase determinations until a
similar analysis can be conducted with limb data for comparison. 
Phase relations described by Ulrich (1996) may not 
apply in sunspots where the fluxtube model is inappropriate.  

%The relationship between the observed magnetogram signal and 
%B field fluctuations is yet to be determined.   
Temperature, density and
corresponding opacity variations may contribute to B amplitudes and 
phases through crosstalk.   
Although the interpretation of the signal variation is complicated by
the effects of crosstalk, the crosstalk is nonetheless of solar origin 
so that the spectral power density maps provide 
a real measure of MHD oscillations.  The distinct spatial distribution 
of magnetogram power in the three frequency regimes can assist 
in region selection of future MHD wave searches as well as 
numerical comparison of amplitude and phase measurements from 
other datasets.

\clearpage

\figcaption[]{Context images and spatial distributions of power are shown.  
Rows from bottom:  context images, 0.5-1.0, 3.0-3.5 and 5.5-6.0 mHz 
frequencies.  Columns from left contain B, I$_c$ and $v$ data.  
Grayscales: black-max, white-min except the I$_c$ context image is 
reversed. Context images from left to right have maximum values of
2057 G, 3436, 1.3 km/s and minimum values of -876 G, 658, -1.3 km/s.  
The maximum values of power from lower to upper plots are
B: 1000, 220, 178 normalized power units, 
I$_c$: 10445, 1690, 280, 
and $v$: 53$\times$10$^3$, 151$\times$10$^3$, 
10$\times$10$^3$ (m/s)$^2$. Minimum power values are
B: 7.2, 8.4, 6.9 normalized power units, 
I$_c$: 10, 3.2, 2.2, and $v$: 153, 1650, 70 (m/s)$^2$.}

\figcaption[]{The logarithm of rms normalized magnetic power is plotted as a 
function of frequency for pixels whose mean flux values are within the 
100-300, 600-800 and 1300-1500 Gauss ranges.  The average standard deviation
values used for signal normalization are 13.0, 14.3, 15.9 G, respectively.}

\figcaption[]{Cross amplitude (left) and phase spectra (right) is shown
 for the 0.5-1.0, 3.0-3.5 and 5.5-6.0 mHz regimes.
Phase data is plotted for all spatial points, regardless of cross amplitude significance, in order to display spatial structure of phases. }

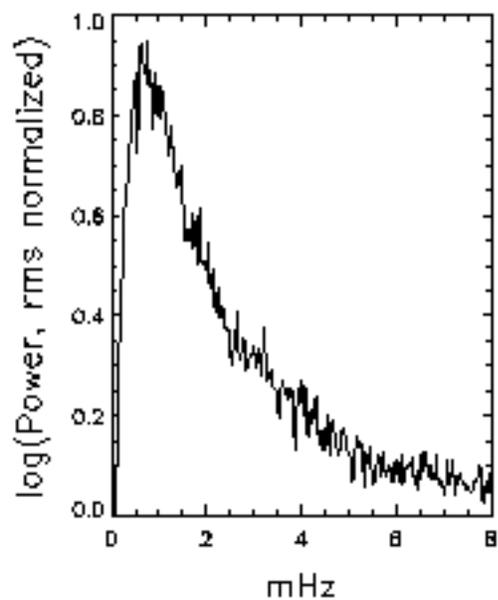
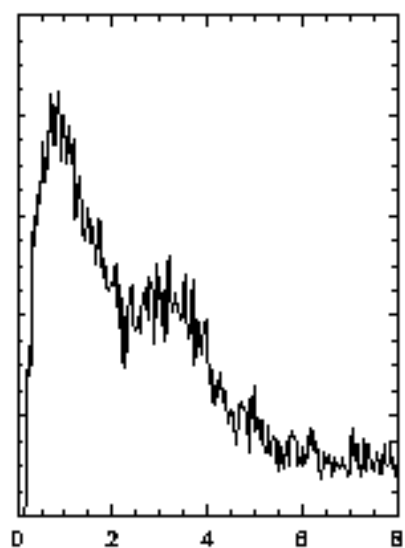
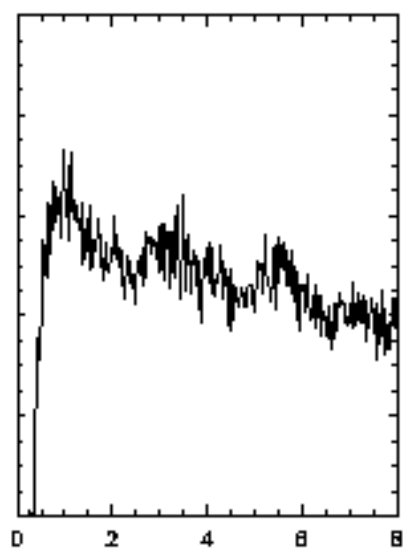
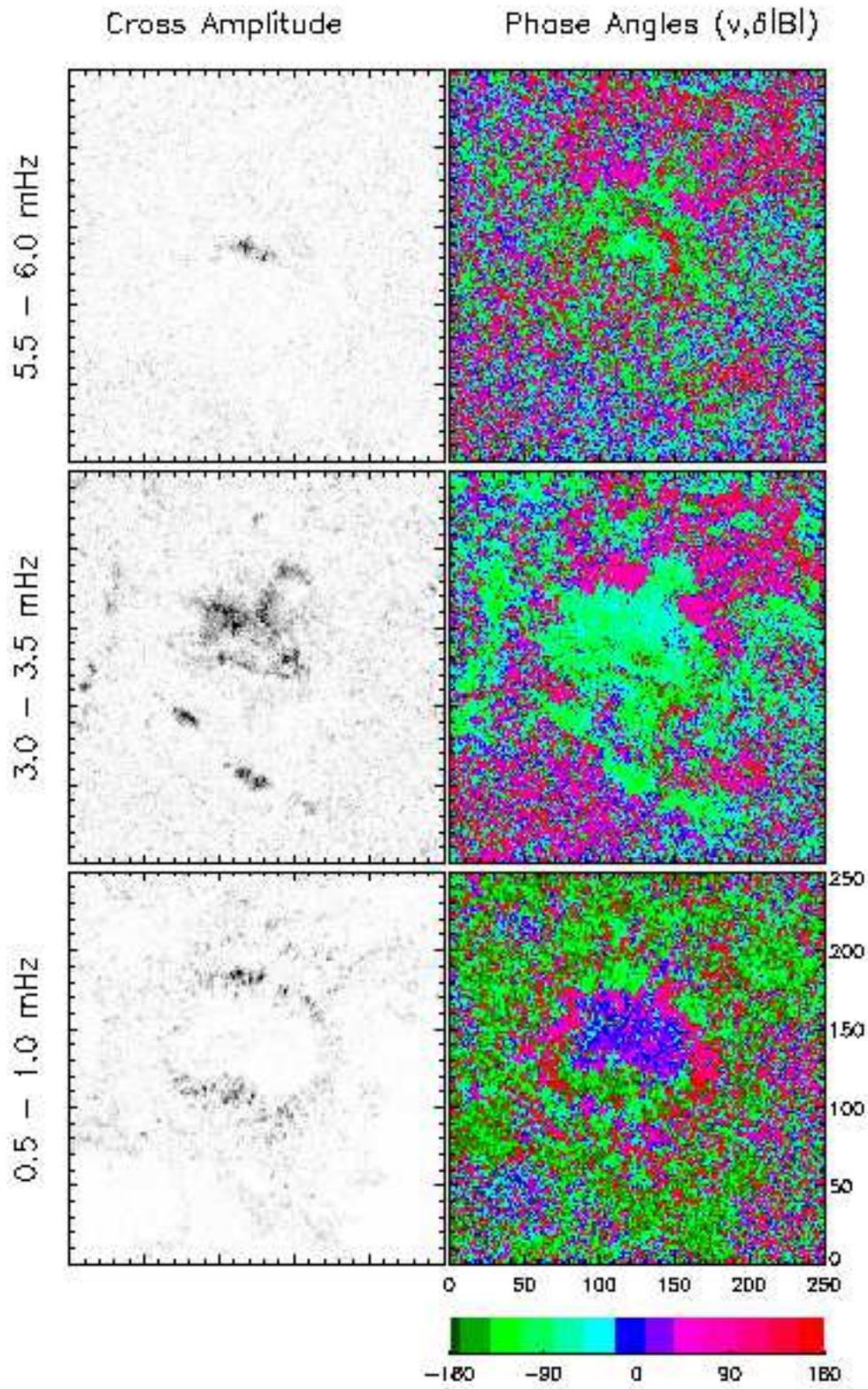
\end{document}